\documentclass[conference]{IEEEtran}
\IEEEoverridecommandlockouts
\usepackage{cite}
\usepackage{amsmath,amssymb,amsfonts}
\usepackage[ruled,linesnumbered]{algorithm2e} 
\usepackage{algpseudocode}
\usepackage{graphicx}
\usepackage{textcomp}
\usepackage{pifont}
\usepackage{xcolor}
\usepackage{booktabs}
\usepackage{subfigure}
\def\BibTeX{{\rm B\kern-.05em{\sc i\kern-.025em b}\kern-.08em
    T\kern-.1667em\lower.7ex\hbox{E}\kern-.125emX}}

\begin{document}
\newcommand{\red}[1]{\textcolor{red}{{#1}}}
\title{BF-Meta: Secure Blockchain-enhanced Privacy-preserving Federated Learning for Metaverse}
\author{
\IEEEauthorblockN{Wenbo Liu}
\IEEEauthorblockA{\textit{The Department of Electrical} \\
\textit{and Electronic Engineering} \\
\textit{The University of Hong Kong}\\
Hong Kong, China \\
u3605164@connect.hku.hk}
\and
\IEEEauthorblockN{Handi Chen}
\IEEEauthorblockA{\textit{The Department of Electrical} \\
\textit{and Electronic Engineering} \\
\textit{The University of Hong Kong}\\
Hong Kong, China \\
hdchen@connect.hku.hk}
\and
\IEEEauthorblockN{Edith C.H. Ngai*}
\IEEEauthorblockA{\textit{The Department of Electrical} \\
\textit{and Electronic Engineering} \\
\textit{The University of Hong Kong}\\
Hong Kong, China \\
chngai@eee.hku.hk}
}

\maketitle

\begin{abstract}
The metaverse, emerging as a revolutionary platform for social and economic activities, provides various virtual services while posing security and privacy challenges. Wearable devices serve as bridges between the real world and the metaverse. To provide intelligent services without revealing users' privacy in the metaverse, leveraging federated learning (FL) to train models on local wearable devices is a promising solution.
However, centralized model aggregation in traditional FL may suffer from external attacks, resulting in a single point of failure. Furthermore, the absence of incentive mechanisms may weaken users' participation during FL training, leading to degraded performance of the trained model and reduced quality of intelligent services. In this paper, we propose BF-Meta, a secure blockchain-empowered FL framework with decentralized model aggregation, to mitigate the negative influence of malicious users and provide secure virtual services in the metaverse. In addition, we design an incentive mechanism to give feedback to users based on their behaviors. Experiments conducted on five datasets demonstrate the effectiveness and applicability of BF-Meta.
\end{abstract}


\renewcommand{\thefootnote}{}
\footnotetext{*Corresponding author: Edith C.H. Ngai.}

\section{Introduction}
The metaverse integrates virtual reality \cite{wohlgenannt2020virtual}, artificial intelligence \cite{huynh2023artificial}, 3D constructions \cite{wu2024cl}, and other advanced digital techniques to construct a worldwide virtual community. The development of advanced applications for the metaverse, such as virtual games, virtual socialization, and remote healthcare, is transforming the way people engage with the world. Leveraging machine learning to provide intelligent personalization and optimize feedback significantly improves the quality of services (QoS) in metaverse applications. 
Nevertheless, the training of machine learning models relies heavily on collecting user data, which introduces significant challenges for users' privacy and security \cite{otoum2024machine}. Federated learning (FL) provides a distributed training scheme for intelligent services in the metaverse while preserving personal privacy \cite{zhang2021survey}.

However, assuming all participants are trustworthy in FL is unrealistic. Some malicious users may exploit the training process for personal gain or disrupt the system training performance \cite{RobustFL_2022, 10018261_TBD}. Additionally, when users enjoy intelligent services in the metaverse, wearable devices continuously collect personal information to train local models. The curious server can easily infer training data distribution and obtain users' personal information with multiple updated local models. Furthermore, once the server is attacked, the risk of privacy leakage increases dramatically, potentially compromising the QoS and leading to system failure.
Applying blockchain to FL can mitigate multiple attacks during model training while ensuring secure data storage \cite{xu2024exploring, mahammad2023scalable}. Through the design of incentive and consensus mechanisms, blockchains also encourage and monitor users to participate in FL honestly \cite{zheng2018blockchain, 9580624, dengm2f} to ensure the QoS in the metaverse.

Early blockchain-empowered FL systems are designed to record users' credits in blocks. Tian \textit{et al.} \cite{tian2023blockchain} propose a credit investigation solution to identify potentially malicious users by establishing a real-virtual combined credit system. The authors implement a real-name registration mechanism to transfer credits from the real world, incentivizing honest participation in FL.
Kang \textit{et al.} \cite{kang2022blockchain} apply blockchain with a designed incentive mechanism in FL to encourage users' participation.
Gupta \textit{et al.} \cite{gupta2023metaverse} investigate the deployment of blockchain for user authentication and the documentation of illegal behaviors in the metaverse. The authors propose recording illegal behaviors on the blockchain and adjusting user reputation based on these records. However, these studies lack a mechanism to verify uploaded models before aggregation to resist poisoned models and ensure the robustness of training, which is significant for models' performance.
Li \textit{et al.} \cite{li2020blockchain} propose a verification mechanism to ensure the performance of local model parameters. However, this verification process is time-consuming. A more efficient detection method needs to be developed for time-sensitive services in the metaverse. 

To address these challenges, this study proposes BF-Meta, a novel blockchain-empowered FL framework with decentralized model aggregation, to detect malicious models before aggregation. Furthermore, we introduce a reputation-based incentive mechanism to encourage users' honest updates during training and enhance the security of services in the metaverse.  
The main contributions are listed as follows:
\begin{itemize}
    \item We propose BF-Meta, a secure blockchain-empowered FL framework to collect data from physical wearable devices and provide services in the metaverse. A consensus mechanism in BF-Meta is designed to verify FL models and provide secure model storage to ensure the security of training even with potential malicious clients.
    \item A reputation-based incentive mechanism is designed to monitor users' behaviors, detect abnormal behaviors, and exclude malicious users during FL training.
    \item We demonstrate the generalisability of BF-Meta and the stability of the framework's performance on five datasets. The sufficient experimental results prove the effectiveness of our proposed framework.
\end{itemize}

The rest of the paper is organized as follows: Section II presents the blockchain-empowered FL framework for the metaverse. Section III details the design of the incentive mechanism. Section IV analyzes the security and experiment performance of BF-Meta. Section V concludes the paper.


\section{Blockchain-empowered FL Framework for metaverse}
In this section, we describe the system architecture and workflow of the proposed BF-Meta. 

\subsection{System Overview}

 \begin{figure}[!t]
    \centering
    \includegraphics[width=\linewidth]{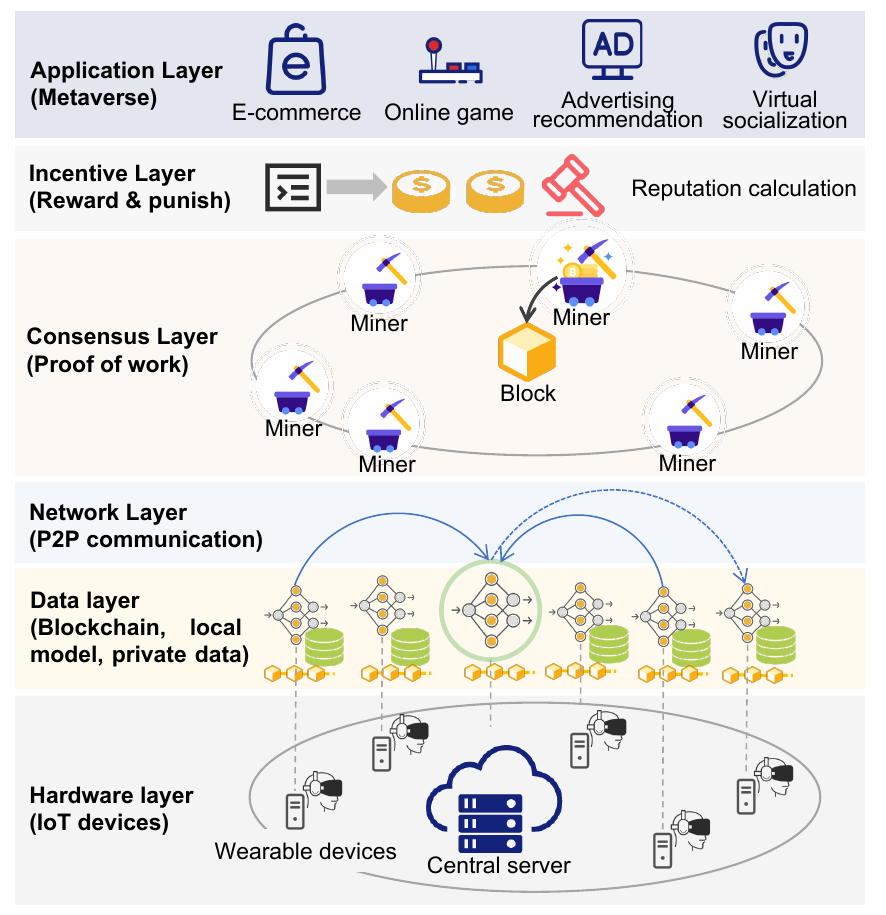}
    \caption{System overview of BF-Meta.}
    \label{fig:sys}
\end{figure}

Fig. \ref{fig:sys} shows the architecture of BF-Meta consisting of six layers from bottom to top: hardware, data, network, consensus, incentive, and application layers. 
\subsubsection{Hardware Layer}
The hardware layer includes multiple wearable devices and a central server. Users' data are collected when using wearable devices and for local training. The server is responsible for verifying and aggregating local models to provide applications. 
\subsubsection{Data Layer}
In the data layer, the stored data includes on-chain and off-chain storage. The collected user information, local models, and blockchains are stored on the respective users' wearable devices. The central server stores models from all clients for verification and aggregation and synchronizes them on the blockchain. The encrypted hash values of the trained model parameters are uploaded and stored on the blockchain along with a digital signature. 
\subsubsection{Network Layer}
In the network layer, peer-to-peer (P2P) communication is leveraged to transmit models between devices. Each update of the blockchain is synchronized through broadcasting.
\subsubsection{Consensus Layer}
Consensus mechanisms prevent any single party or a small group of parties from controlling the entire system. To ensure data interaction security, miners in the consensus employ proof of worklayer. We also utilize an accuracy-based verification mechanism to enhance proof of work consensus and prevent poisoning.
\subsubsection{Incentive Layer}
The incentive layer encourages clients to invest their computation resources and monitor their behaviors during training. BF-Meta utilizes reputation in the metaverse to reward and punish clients according to their behaviors. 
\subsubsection{Application Layer}
After completing the training, the models in BF-Meta can be utilized to provide various services in the metaverse, such as e-commerce, online games, advertising recommendations, and virtual socialization.

\subsection{Threat Models}

In the metaverse, communities may involve thousands of users worldwide. We consider that all users are semi-trusted and may upload stale parameters without training to reduce the computation cost for personal gains. These users are viewed as ``lazy" users. Repeated stale parameters may reduce the accuracy of the global model and hinder training convergence. Therefore, it is necessary to detect lazy participants to eliminate low-quality model parameters from model aggregation. To avoid the detection of hash values stored in blocks, the malicious clients may poison the model by manipulating historical hash values.



\subsection{Latency Calculation}
\begin{figure}[!t]
    \centering
    \includegraphics[width=1\linewidth]{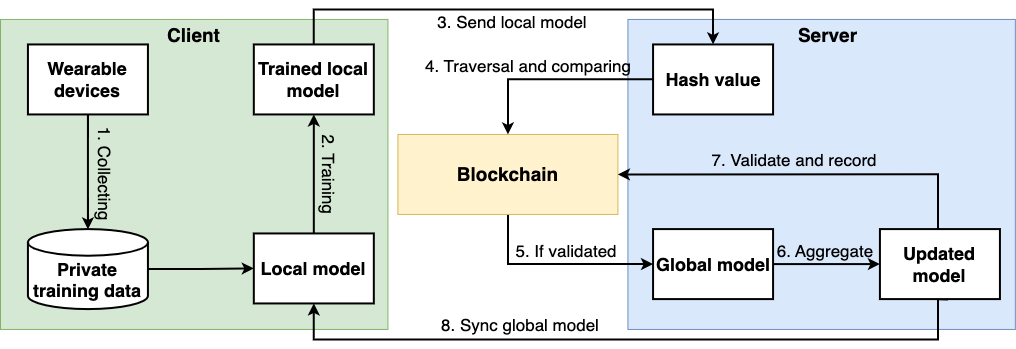}
    \caption{The workflow of BF-Meta. The left and right parts indicate the workflow of the client and the server sides, respectively.}
    \label{fig:process}
\end{figure}

The latency of BF-Meta consists of two parts: system latency $t_{c}$ and blockchain latency $t_{b}$. The system latency $t_{c}$ represents the total time consumption of FL until convergence, consisting of transmission latency between a client and the server (steps 2, 3, and 8 in Fig. \ref{fig:process}), which may be affected by internet conditions and device hardware status. Therefore, the system latency can be represented as: 
\begin{equation}
    t_c=n(t_{fl}+t_{c\rightarrow s}+t_{s\rightarrow c}),
\end{equation}
where $t_{fl}$, $t_{c\rightarrow s}$ and $t_{s\rightarrow c}$ represent the latency of FL training, the transmission latency from client to server, and the transmission latency from server to client in one communication rounds, respectively. $n$ indicates the number of communication rounds required for convergence.
The blockchain latency $t_{b}$ denotes the time consumption related to blockchain verification, which includes the latency of block generation $t_{bg}$ (step 4 in Fig. \ref{fig:process}), block consensus $t_{bv}$ (steps 5 and 7 in Fig. \ref{fig:process}), and blockchain synchronous $t_{bs}$ (step 8 in Fig. \ref{fig:process}). The latency can be calculated by equation (\ref{equ:bc}) as follows:
\begin{equation}\label{equ:bc}
    t_b=t_{bg}+t_{bv}+t_{bs}.
\end{equation}
Herein, the latency of block consensus may be impacted by the difficulty of the consensus mechanism.

\subsection{Secure FL Training in BF-Meta}

\begin{algorithm}
    \label{alg1}
    \caption{Training process in BF-Meta.}
    \textit{/* $Update(k, \omega_k^t)$ on client $k$*/}\\
    Collect training datasets from wearable devices\;
    Synchronous updated blockchain\;
    Obtain the global model\;
    \ForEach{epoch $e$}{
    \ForEach{batch $b$}{
    $\omega_k^t\xleftarrow{}\omega_k^{t-1} - \eta\nabla\delta(\omega_k^{t-1};b)$\;
    }
    }
    \Return the local model to the server for verification.\\
    \textit{/* $Verify(\omega_k^t)$ on server */}\\
    Calculate the hash value of $\omega_k^t$\ ($Hash(\omega_k^t)$);
    \If{$Hash(\omega_k^t)$ is not duplicated}{
        $\beta_k=0$\;
        \Return send validated model $\omega_k^t$ to miner\;
    }
    \textbf{else} $\beta_k=1$\;
    \Return error message to client $k$.\\
    \textit{/* Block storing $Hash^b(\omega_k)$ will be uploaded to the blockchain via proof of work. */}\\
    \textit{/* $Aggregate(\mathcal{W})$ on server */}\\
    Synchronous updated blockchain\;
    \ForEach{model $\omega_k^t$ in model list $\mathcal{W}$}{
        \If{$Hash^b(\omega_k^t)=Hash(\omega_k^t)$ }{
            $\alpha_k=0$\;
        }
    }
    \textbf{else} $\alpha_k=1$\;
    $\omega^{t+1}\xleftarrow{}\sum_{k=1}^{K}(1-\alpha_k)(1-\beta_k)\frac{n_k}{n}\omega^{t+1}_k$\;
    Send model $\omega^{t+1}_k$ to miner for uploading to blockchain\;
    \Return $\omega^{t+1}_k$ to all clients.
\end{algorithm}



The operations of this blockchain-empowered FL system can be divided into eight main steps, as shown in Fig. \ref{fig:process}. The detailed procedure is explained as follows:
\begin{enumerate}
    \item Users' historical records and preference information are collected as a private training dataset. 
    \item Users train the local models based on the private dataset with the received models from the last synchronization. If it is the first round, the initial global model is used. 
    \item Clients send their trained local model to the miner deployed with blockchain service. The transmitted data are encrypted by the hash function. Miners firstly verify the received blocks with the proof of work consensus mechanism by calculating a Nonce value that makes the hash value of the new block less than a target value and then upload the new block to the blockchain in sequence. 
    \item The server verifies the validity of the received blocks by traversing and comparing the hash values stored in the blockchain. 
    \item After verifying, qualified local models are downloaded to the server for model aggregation 
    \item The server aggregates local models to generate the global model for FL.
    \item The updated global model is validated and uploaded to the blockchain. 
    \item The latest status of the blockchain is synchronized to all participants. By downloading updated models from the latest blockchain, clients can update their local models for the next round of training. 
\end{enumerate}
The step numbers correspond to numbers in Fig. \ref{fig:process}, repeating steps 2-8 until the model converges.

\section{Reputation-based Incentive Mechanism}
To ensure the performance of training, an incentive mechanism plays a crucial role in ensuring all clients are motivated to participate in training honestly. The incentive mechanism involves reward and punishment schemes to monitor clients’ behaviors. This section explains the design of client monitoring and incentive schemes in the proposed BF-Meta. 

\subsection{Client Monitoring Scheme}
To detect and record malicious behaviors in FL, a blockchain-enabled incentive mechanism is leveraged to assist the server in model selection for aggregation. 
In BF-Meta, each client is assigned a unique address for recording and identifying both clients and their respective models. Once the local parameters are received, the server traverses the blockchain to retrieve the latest model parameters uploaded by the same client from the blockchain with the unique address. After that, the server compares the hash value of the received models with the hash value of the previous models. If they match, the client will be detected as a lazy client and added to a blacklist. 


To avoid false positive detections of lazy clients, a threshold for model accuracy is introduced. The threshold is set as the average accuracy of all received models in each aggregation. Upon receiving the local models, the server uses a partial test dataset to evaluate their reliability. If the assessed accuracy is lower than the accuracy threshold, the server logs this result and initiates a historical query on the blockchain to investigate potential causes.
Any lazy and low-accuracy models are categorized as ``low-quality" and will be excluded from model aggregation. Furthermore, the server computes the hash value of the received model and queries the blockchain to compare it with the historical hash values stored to detect data falsification. Once data falsification or repeated model is detected, the data falsification indicator ($\alpha$) or lazy client indicator ($\beta$) will be set as 1, respectively. The corresponding punishment will be reflected in the client's reputation, as mentioned in Section IIIB.

\subsection{Incentive Mechanism}
This section introduces the designed incentive mechanism in BF-Meta to motivate clients by detecting and punishing malicious behaviors through the client monitoring scheme and rewarding honest clients accordingly. A typical incentive mechanism \cite{zhao2019mobile} is to apply all the past performances to determine the assessment weights. From a long-term perspective, the historical performance of all clients in the vast metaverse may exhibit similarities due to analogous variations in model accuracy over time. Applying all records to determine weights can be challenging, as it may not sufficiently increase the gaps among clients' reputations and effectively encourage their participation in FL. Therefore, we introduce a time-varying factor for the reputation mechanism, where the performance of the last model aggregation will directly determine the current reputation.

Specifically, we define two reputation factors to evaluate the weights of received models: model quality and data quantity. These two factors denote the accuracy of the local model and the size of the training datasets, respectively. The server sampled the test dataset to test the model's accuracy and evaluate the model's quality. The server transforms the test accuracy into data quality weight ($\omega_{k}$). The model quality factor is calculated by: 
\begin{equation}
    \omega_{k} = \frac{\theta_{k} - \theta_{min} }{\theta_{max} - \theta_{min}},
\end{equation}
where $\omega _{k}$ is the model quality factor, and $\theta_{k}$ denotes the test accuracy of client $k$. $\theta_{min}$ and $\theta_{max}$ represent the minimum and maximum values of test accuracy among clients selected for aggregation.

Data quantity factor ($\varphi_{k}$) is calculated depending on the size of data the client collects and utilizes. The data quantity factor can be obtained by:
\begin{equation}
    \varphi_{k} = \frac{\L_{k} - \L_{min} }{\L_{max} - \L_{min}}, 
\end{equation}
where $\varphi_{k}$ denotes the data quantity factor, and $\L_{k}$ denotes the data size of client k. $\L_{min}$ and $\L_{max}$ represent the minimum and maximum values of data size among clients in FL, respectively.

Based on these two factors, the incentive mechanism is operated based on the basic reputation ($R_{basic}$), data quality ($R_{quality}$), and the data quantity ($R_{quantity}$) to calculate the rewarded or punished reputation in a round of model aggregation. The reputation is calculated as follows:
\begin{align}
    \begin{split}
        R_{k}^{t} =& (1-\beta_k)(1-\frac{\alpha_k}{2})(R_{k}^{t-1}+R_{basic}+R_{quantity}\omega_{k}^t+\\
        &R_{quality}\varphi_{k}^{t-1}),
    \end{split}
\end{align}
where $\omega_{k}^{t}$ and $\varphi_{k}^{t-1}$ are factors to indicate the data quantity and model quality, respectively.

In BF-Meta, the selection of clients for the following round aggregation is closely tied to their reputation value. Specifically, normalization is employed to calculate probabilities based on reputation values. Reputation values are transformed into probabilities ($0\sim1$). Clients with higher reputations have a higher probability of being selected for aggregation in the following round of training. Two reputation factors, model quality and data quantity, determine the reputation feedbacked to clients in each round. The reputation updates will be repeated until the training is finished.

\section{Security Analysis and Experimental Study}
In this section, we first discuss several implementation cases. Then, we evaluate the performance of BF-Meta by security analysis and experimental study. 

\subsection{Implementation Discussion}
BF-Meta can be implemented on the metaverse to provide intelligent services. 

\subsubsection{Recommendation System}
Metaverse is a virtual mapping of the real world. Personalized recommendation services are necessary to attract and retain users. To achieve this target, the service providers need to collect users' browsing history and behaviors to train a model for offering personalized service recommendations. BF-Meta, with the incentive mechanism, can be implemented as a secure distributed training framework to train local models for personalized service without revealing data privacy and encourage operators to share their user information.

\subsubsection{Credit Investigation}
Credit investigation is crucial for assessing the risk level of users in the metaverse by reviewing their credit information from the real world. In this case, BF-Meta can be leveraged to securely train an evaluating model to evaluate the risk of user transactions to guarantee the security of the metaverse. Security of BF-Meta can effectively protect the users' information and training process data from falsification and attack.

\subsubsection{Decentralized Social Platform}
BF-Meta can also be applied to the architecture of social platforms in the metaverse. By collecting users' historical behaviors and information, operators can train local models for like-minded people's recommendations. Users with similar models can be recommended as friends.

\subsection{Security Analysis}
BF-Meta can resist several attacks. We analyze the security from the following attacks.

\subsubsection{Model Poisoning Attacks}
In FL, model poisoning attacks aim to corrupt the shared model by falsifying the transmitting model parameter or continuously uploading untrained model parameters. With BF-Meta, every model update is verified by comparing the hash values between the received model and the previous hash value stored on the blockchain of the same client so that the repeatedly transmitted model can be detected and that client will be blacklisted. Moreover, if a malicious client poisons his model when updating. The accuracy of his model will be lower than that of other honest clients'. Once a low-quality model is detected, the reputation penalty will reduce its probability of being selected in the following rounds, thereby forfeiting the model reward.

\subsubsection{Sybil Attacks}
In a Sybil attack, an attacker creates multiple fake identities to gain a disproportionate influence over the network. BF-Meta can validate each participant's identity when receiving the model. By comparing the signature with the authorization list of the blockchain, the timestamp of model uploading, and model quality, the server will identify the fake identity. If the signature of the attacker is not in the authorization list of the blockchain, it will be detected as a malicious node and blacklisted.

\subsubsection{Replay Attacks}
In a replay attack, an attacker intercepts a valid data transmission and retransmits it to create unauthorized effects. In this case, the speed and efficiency of training can be reduced by channel blocking and training obstruction.
BF-Meta's use of timestamps and sequence numbers in transaction records helps prevent replay attacks by ensuring that each transaction is uniquely identified and cannot be reused. When receiving model parameters, BF-Meta compares the hash values and timestamps between the current received model and the previous ones of the same client. Once the replay attacks are detected, the client will be blacklisted with its reward forfeited.


\subsection{Experimental Study}

\subsubsection{Experiment Settings}

The experiments are built with Python 3.8.0 and Pytorch running on CPU 12th Gen Intel(R) Core (TM) i7-12700H to establish BF-Meta and use Web3 to achieve the interaction between FL training and the blockchain system. The settings of the experiment are listed in Table \ref{tab:nota}.
We consider a metaverse platform with $K = 30$ users participating in FL with private keys. A selection rate $\eta$ is introduced to represent the chosen clients for model aggregation. Moreover, we set the target accuracy equal to $80\%$ as the criterion for the termination of FL training. Based on the clients' behaviors, participants are divided into two categories: normal clients and lazy clients. Normal clients participate in training honestly, while lazy clients become ``free-riders" in FL as mentioned in the threat models. 
We implement BF-Meta by Pytorch and a blockchain platform named Ganache based on five public datasets. The datasets and the corresponding models are detailed as follows: 
\begin{itemize}
    \item MovieLens: we use a deep neural network (DNN) model with 5 hidden layers, and the numbers of neurons in each layer are 20, 30, 50, 30, and 20, respectively;
    \item Wisconsin Breast Cancer Dataset: we use a DNN model with 6 hidden layers, and the numbers of neurons in each layer are 10, 30, 40, 40, 30, and 10, respectively;
    \item Wheat Seed Dataset: we use DNN models with 4 hidden layers, and the numbers of neurons in each layer are 10, 30, 30, and 10, respectively;
    \item Pima Indian Diabetes Dataset: we use a DNN model with 6 hidden layers, and the numbers of neurons in each layer are 20, 30, 50, 50, 30, and 20, respectively; 
    \item MNIST: we use a CNN model with 4 convolutional layers, 4 pooling layers, and 3 full-connecting layers whose neuron numbers in each are 256, 128, and 10, respectively.
\end{itemize} 


\begin{table}[!t]
    \centering
    \caption{Default experimental settings.}
    \label{tab:nota}
    \begin{tabular}{cc}
    \toprule
     \textbf{Parameters} & \textbf{Setting} \\
     \midrule
     Target accuracy (Acc)	& 80\%\\
Total clients in FL (K)	&30\\
Malicious rate 	& [3\%, 10\%, 30\%, 50\%]\\
Client rate for aggregation &50\%\\
\bottomrule
    \end{tabular}    
\end{table}

\subsubsection{Experimental Evaluation}
\begin{table*}[htbp]
    \centering
    \caption{Latency over different training datasets.}
    \label{tab:latency}
    \begin{tabular}{ccccc}
    \toprule
        \textbf{Datasets} & \textbf{Accuracy} & \textbf{Average training latency} & \textbf{Average blockchain latency} & \textbf{Total latency}\\
        \midrule
        \textit{MovieLens} & 81.20 \% & 35.5 s & 3.027 s & 11min 24 s \\
        \textit{Wisconsin Breast Cancer Dataset} & 80.74 \% & 40.43 s & 2.591 s & 12 min 11 s \\
        \textit{Wheat Seed Data} & 81.63 \%	& 20.4 s	& 2.654 s &	6 min 9 s\\
        \textit{Pima Indian Diabetes Dataset} &80.17 \%	&47.71 s&	2.431 s&	12 min 30s \\
        \textit{MNIST} &83.72 \%	& 1 min 46s &	3.254 s	&36 min 25s\\
        \bottomrule
    \end{tabular}
\end{table*}

We simulate the latency on the mentioned datasets during FL training to evaluate the stability of BF-Meta and analyze the relationship between the accuracy and malicious user ratio in clients to evaluate the performance of the detection function. 

\begin{figure}
    \centering
    \includegraphics[width=0.9\linewidth]{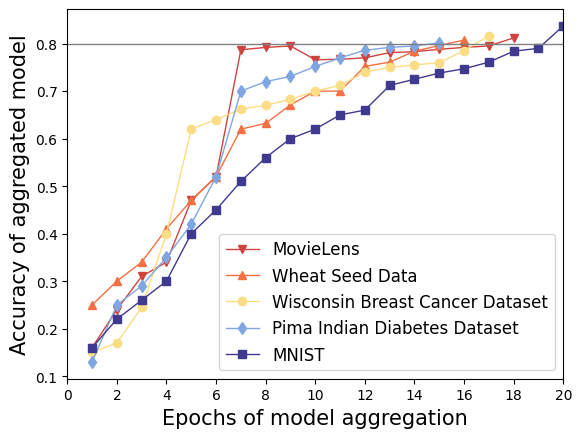}
    \caption{Accuracy of BF-Meta on five datasets.}
    \label{fig:acc}
\end{figure}

Table \ref{tab:latency} shows the latency of BF-Meta on different datasets. The target accuracy is set as 80\%. We evaluate the system, blockchain, and total latency as explained in section II. Fig. \ref{fig:acc} shows the variation trends of the aggregated model in each dataset in terms of model aggregation epoch. Even if there are obvious differentiations in total latency and system latency among these datasets, average blockchain latency remains stable. In Table \ref{tab:latency}, both total latency and average system latency experiments on each dataset are different, distributing from 6 min to 36 min in total latency and 20 seconds to nearly 2 min in system latency. The difference is caused by different model structures used for distinct datasets. An obvious difference occurs between MNIST, a graphic dataset, and other digital datasets because the convolutional neural network (CNN) requires more time to load, process images, and calculate gradients, while the time consumption for the same operations in the feed-forward neural network is lower. However, the latency of BF-Meta on different datasets are similar since the blockchain latency is only influenced by the efficiency of the blockchain committee, miners, and the network condition in the FL training process. Table \ref{tab:latency} and Fig \ref{fig:acc} demonstrate that the BF-Meta we proposed has excellent generalization capabilities across different training tasks. 

\begin{figure}[ht]
    \centering
    \subfigure[FedAvg.]
    {
        \includegraphics[width=0.46\linewidth]{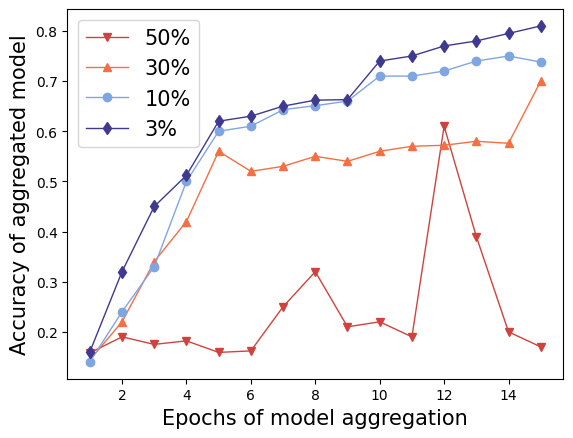}
        \label{exp:1a}
    }
    \subfigure[BF-Meta.]
    {
        \includegraphics[width=0.46\linewidth]{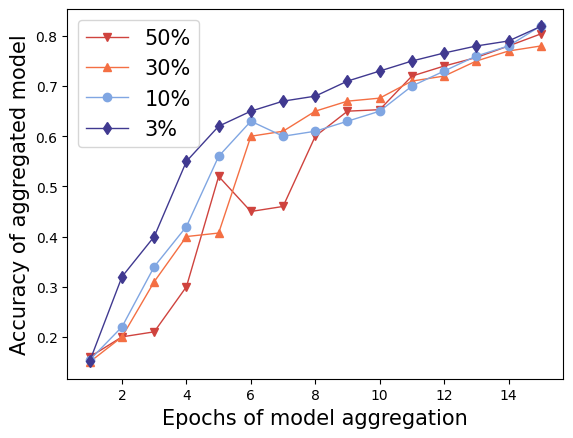}
        \label{exp:1b}
    }
    \caption{Accuracy over epochs of FedAvg and BF-Meta.}
    \label{exp:1}
\end{figure}

\begin{figure}[ht]
    \centering
    \subfigure[3\% malicious clients.]
    {
        \includegraphics[width=0.46\linewidth]{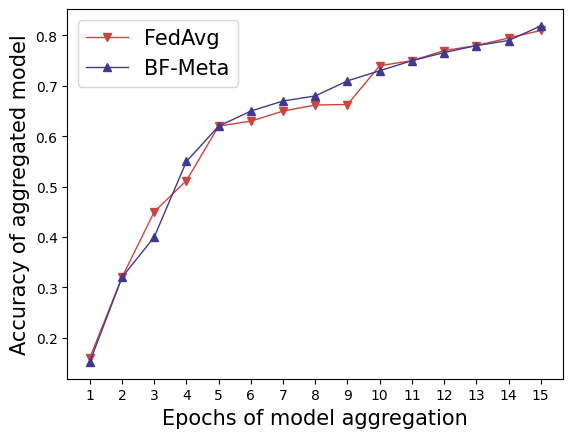}
        \label{exp:2a}
    }
    \subfigure[10\% malicious clients]
    {
        \includegraphics[width=0.46\linewidth]{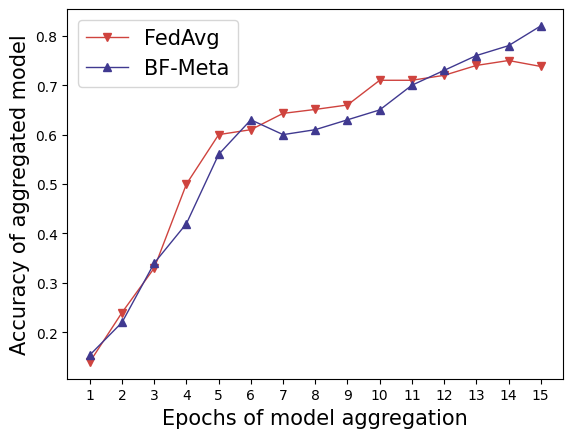}
        \label{exp:2b}
    }
    \subfigure[30\% malicious clients.]
    {
        \includegraphics[width=0.46\linewidth]{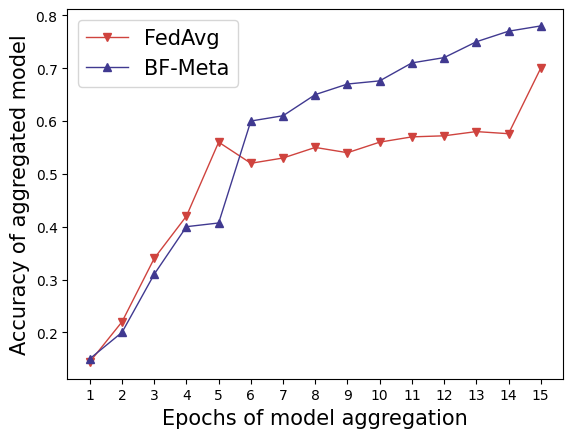}
        \label{exp:2c}
    }
    \subfigure[50\% malicious clients]
    {
        \includegraphics[width=0.46\linewidth]{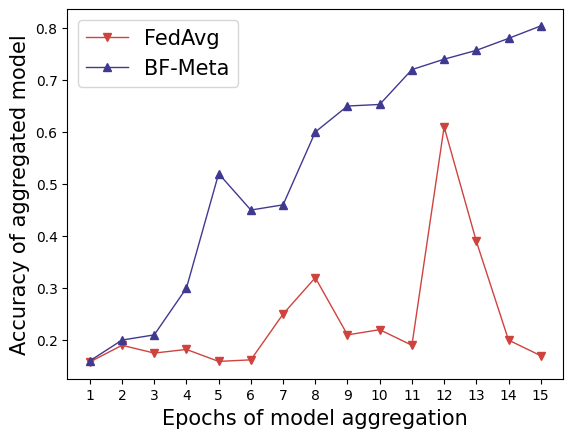}
        \label{exp:2d}
    }
    \caption{Comparison of accuracy of FedAvg and BF-Meta with distinct rates of malicious users.}
    \label{exp:2}
\end{figure}


Fig. \ref{exp:1} shows the comparison between the model accuracy of FedAvg and BF-Meta with distinct malicious client rates in 30 participants and a fixed selection rate of 50\% in FL. In Fig. \ref{exp:1a} and Fig. \ref{exp:1b}, 
as the rate of malicious clients increases, fluctuations in the aggregated model accuracy intensify in the absence of lazy client detection. Conversely, the accuracy fluctuation of BF-Meta is reduced. The reason is that client selection during model aggregation is completely random. It means that each client has the same possibility to be selected in model aggregation in traditional FL. The aggregated model accuracy will be impacted when malicious clients are selected for aggregation. Moreover, more malicious clients are likely to be selected in model aggregation with the increasing number of malicious clients amount, so the fluctuation in 50\% malicious clients is more dramatic than that in other rates. 

To detail the difference, Fig \ref{exp:2} illustrates the accuracy comparison of distinct malicious client rates. As shown in Fig. \ref{exp:2a} and Fig. \ref{exp:2b}, the accuracy of FedAvg is slightly lower than that of BF-Meta. As shown in Fig. \ref{exp:2c} and Fig. \ref{exp:2d}, with the increasing malicious rate, the superiority in accuracy becomes more obvious. In Fig. \ref{exp:2c}, the accuracy of BF-Meta is 12\% higher than FedAvg since BF-Meta detects and defends against attacks by lazy clients. As shown in Fig. \ref{exp:2d}, the accuracy gap between BF-Meta and FedAvg is much more obvious. The model accuracy of BF-Meta increases stably along the training epoch, while FedAvg cannot converge at all due to lazy clients. 
The experimental results demonstrate that malicious clients occupy a large proportion of clients in FL so that more bad parameters are uploaded in model aggregation. For BF-Meta with the client monitoring scheme, malicious clients are blacklisted once they upload the repeated parameters. Correspondingly, the reputation will be zeroed until the end of FL. For epoch 5 in Fig. \ref{exp:2d}, the decrease in accuracy is caused by the selection of malicious clients in model aggregation. Once lazy clients are detected and blacklisted, the model accuracy increases again. These experimental results demonstrate the client monitoring scheme in BF-Meta can effectively avoid the influence of malicious clients and improve accuracy. 
\begin{figure}[ht]
    \centering
    \subfigure[Initial reputation.]
    {
        \includegraphics[width=1\linewidth]{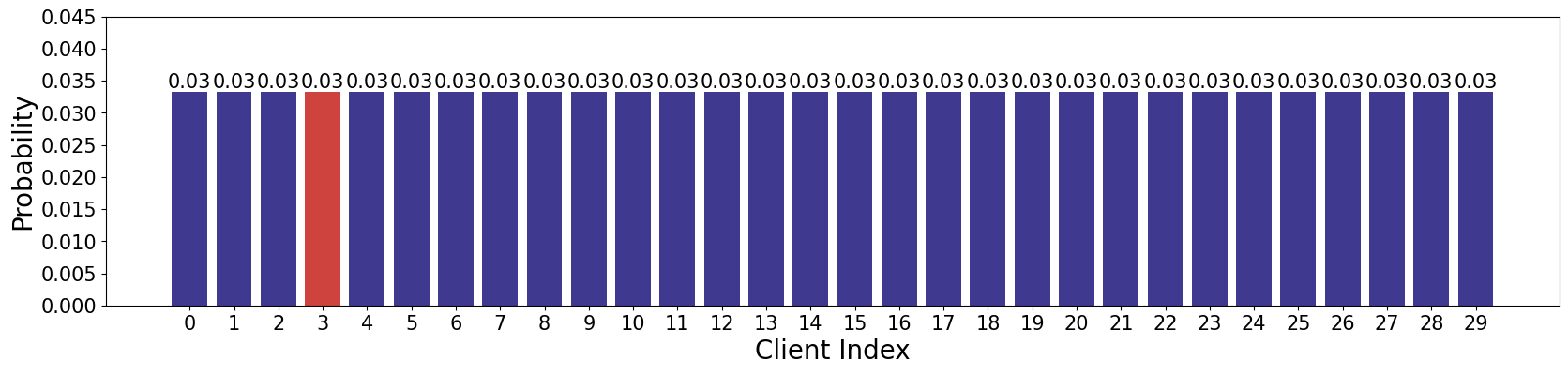}
        \label{exp:30}
    }
    \subfigure[Reputation after one round.]
    {
        \includegraphics[width=1\linewidth]{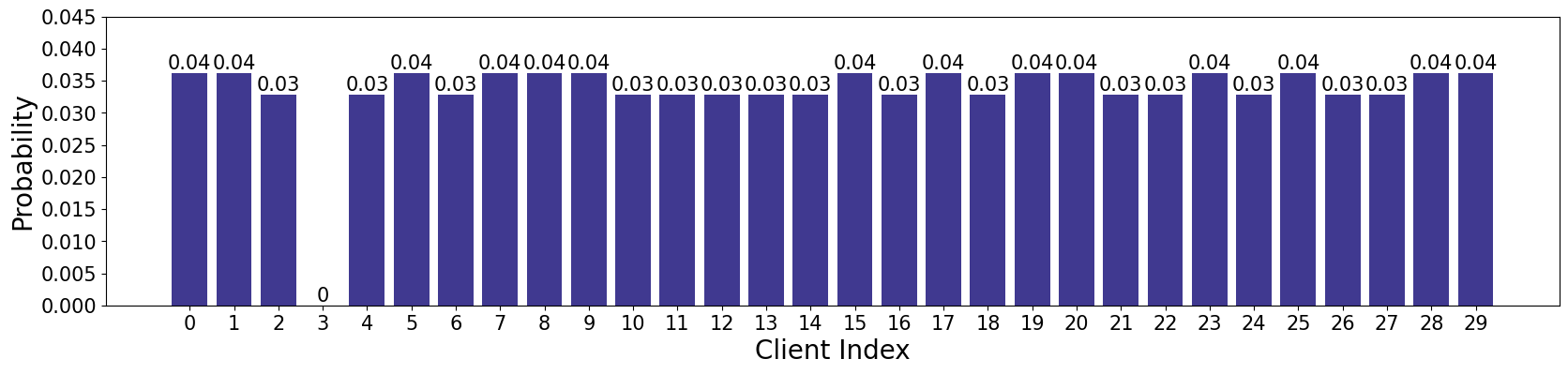}
        \label{exp:31}
    }
    \subfigure[Reputation after two rounds.]
    {
        \includegraphics[width=1\linewidth]{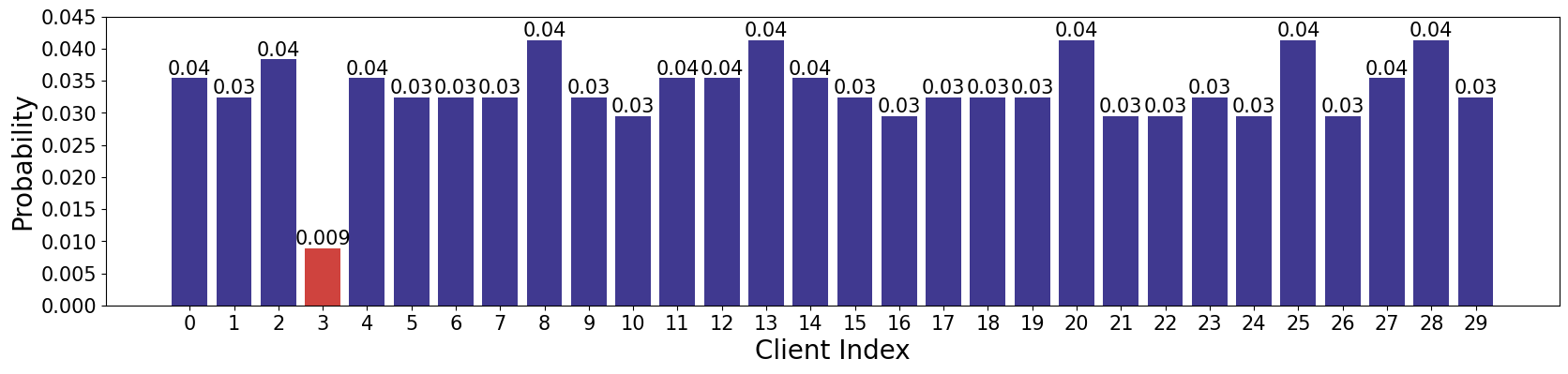}
        \label{exp:32}
    }
    \caption{Variations in client reputation over rounds. Herein, the blue bars represent normal clients while the red bar represents the malicious client (client index = 3).}
    \label{exp:3}
\end{figure}
Fig. \ref{exp:3} illustrates the variation in clients' reputations to evaluate the incentive mechanism. In Fig. \ref{exp:30}, at the beginning of FL, all the clients are given the same reputations. Each client has an equal probability of being selected for model aggregation, even if the malicious client is marked in red (index 3). 
Fig. \ref{exp:31} shows the reputations of attack detection of malicious clients after one round. When detecting the inequality of the hash value of the malicious client (index 3), the incentive mechanism sets its reputation to zero so that the malicious client (index 3) will not be selected at the next model aggregation. In the meantime, the reputation of the normal clients selected in model aggregation increases, so that these clients are more likely to be selected in the next model aggregation than others. In this case, the malicious model parameters can be rejected from model aggregation.
Fig. \ref{exp:32} indicates the reputations after two rounds, where the reputation of the malicious client (index 3) is much lower than others. The minor increase in malicious client (index 3) reputation is caused by the basic reputation in FL. However, the probability of the malicious client (index 3) being selected is much less than the others.

\section{Conclusion}
In this paper, we propose BF-Meta, a secure blockchain-empowered FL framework for the metaverse that incorporates incentive and malicious client monitoring schemes. It can efficiently resist model poisoning attacks and detect lazy clients, thereby ensuring the security of FL training tasks. Furthermore, we introduce a reputation-based incentive mechanism to encourage users to participate honestly in FL training by providing feedback on their behaviors. Overall,  BF-Meta provides a secure framework to support intelligent services in the metaverse without revealing private user information.

\section*{Acknowledgement}
This research was supported by Meta AR/VR Policy Research Fund, HKU-SCF FinTech Academy, and Hong Kong General Research Funds (grant no. 17203320, 17209822).

\bibliography{references}
\bibliographystyle{ieeetr}

\end{document}